\documentclass[pdftex]{article}
\usepackage{icrctc07}

\title{The performance of an idealized large-area array of moderate-sized IACTs}
\shorttitle{The performance of a idealized large-area array of moderate-sized IACTs}
\authors{S. J. Fegan and V. V. Vassiliev}
\shortauthors{S. J. Fegan and V. V. Vassiliev}
\afiliations{Department of Physics and Astronomy, University of California, Los Angeles, CA 90095, USA}
\email{sfegan@astro.ucla.edu}

\abstract{We present simulations of a large array of imaging atmospheric 
Cherenkov telescopes (IACTs), for which the size of the array
footprint is much larger than the size of the Cherenkov lightpool. To
evaluate limitations of the imaging atmospheric Cherenkov technique,
the array is simulated under the assumption of ideal optics, having
infinite resolution of the photon arrival direction, which makes our
conclusions independent of any particular telescope implementation.
The primary characteristics of the array performance, $\gamma$-ray
trigger efficiency, photon energy at the peak of the detection rate,
and angular resolution are calculated as a function of the parameters
of the array: telescope spacing, telescope aperture, and camera
pixelation. We discuss implication of the results for the design of
the next generation ground-based $\gamma$-ray observatory.}

\begin{document}
\maketitle

\section{Introduction}

Next generation ground-based $\gamma$-ray observatories, such as AGIS
and CTA, will likely consist of an array of on the order of 100
telescopes, spread over an area potentially exceeding 1km$^2$ on the
ground. A starting point for understanding the response of such
instruments is to simulate a large array of identical telescopes, with
the parameters of the array covering the range that is reasonable from
the point of view of cost, practicability of construction, and
scientific return.  This constrains the effective light collecting
area of the individual telescopes to the range between 20~m$^2$ to
200~m$^2$, the separation between telescopes to the range of $80$~m
and $217$~m and the pixel size in the camera to the range from 1 arc
minute to 16 arc minutes.

The large-area array will operate in a different regime to the current
generation instruments, VERITAS and H.E.S.S.\  
\cite{REF::WEEKES::AP2002,REF::AHARONIAN::HESS_CRAB_2006}. 
Significantly above 100~GeV the array will have a collecting area
greater than its footprint on the ground, giving it a sensitivity an
order of magnitude larger than VERITAS at high energies. Because the
footprint of such an array is much larger than the characteristic size
of the Cherenkov light-pool from an electromagnetic cascade in the
atmosphere, it will take advantage of the ``cell effect'',
significantly improving its response at energies lower than 100~GeV
\cite{REF::VASSILIEV_HEASTRO_2005}. Finer pixelation of the cameras of
the individual telescopes will further increase the sensitivity,
especially at lower energies. This increase arises from both the
improved angular resolution and increased background rejection. In
this paper we investigate the cell and pixel effects numerically.

\section{Simulations}

Quasi-infinite arrays of hexagonally packed Cherenkov telescopes were
simulated. The arrays are described by two parameters, the separation
between telescopes, $L$, and the effective diameter of each telescope,
$D$, assuming an effective light collection area of
$\frac{\pi}{4}D^2$. We simulate arrays with all combinations of
$L=\{80,91,106,128,160,213\}$~m and $D=\{5,7,10,15\}$~m.

The simulations employed the CORSIKA air-shower package
\cite{REF::HECK::FZKA1998} to generate Cherenkov photons at 3500~m
elevation.  An idealized telescope optical model, with zero dispersion
due to PSF, was assumed. Light losses due to reflections at two
aluminum mirrors and inefficient conversion to photoelectrons at a
standard bi-alkali photocathode were included. The telescopes were
allowed to have different triggering and imaging pixel scales. It was
previously found that the optimal pixel size on the triggering sensor
is between $\sim0.05^\circ$ and $\sim0.25^\circ$, and is relatively
independent of $L$ and $D$ \cite{REF::VASSILIEV_HEASTRO_2005}; we
chose a value of $\sim0.15^\circ$ in this simulation. The trigger
threshold is set to give a 250~Hz per-telescope rate of accidental
triggers from night sky noise in the $7^\circ$ field of view. The
array is triggered when three telescopes detect a simulated
event. Since we have assumed an ideal optical system, we investigated
the effects of PSF and pixelation in the imaging sensor by quantizing
the arrival direction of the Cherenkov photons into pixels once they
reach the telescope. Pixel sizes of $P=\{1,2,4,8,16\}$ arc minutes
were chosen for study. The cleaning procedure, which was based on the
density of photoelectrons in the image, was independently optimized
for each pixel size to achieve the best reconstruction of events.

As the array is assumed to be effectively infinite we take advantage
of the translational symmetry between the hexagonal cells in the array
to simplify the simulations. We use CORSIKA to generate $\gamma$-ray
events that impact within one ``central'' hexagonal cell of the array.
Since in this study we are primarily interested in the response of the
array at low energies, only those telescopes within $~\sim600$ meters
of the central cell were simulated. More distant telescopes do
not contribute to the triggering and reconstruction of $\gamma$-ray
events at low energy.

\section{Results}

\begin{figure*}[th]
\begin{center}
\resizebox{0.99\textwidth}{!}{\includegraphics[height=0.5\textwidth]{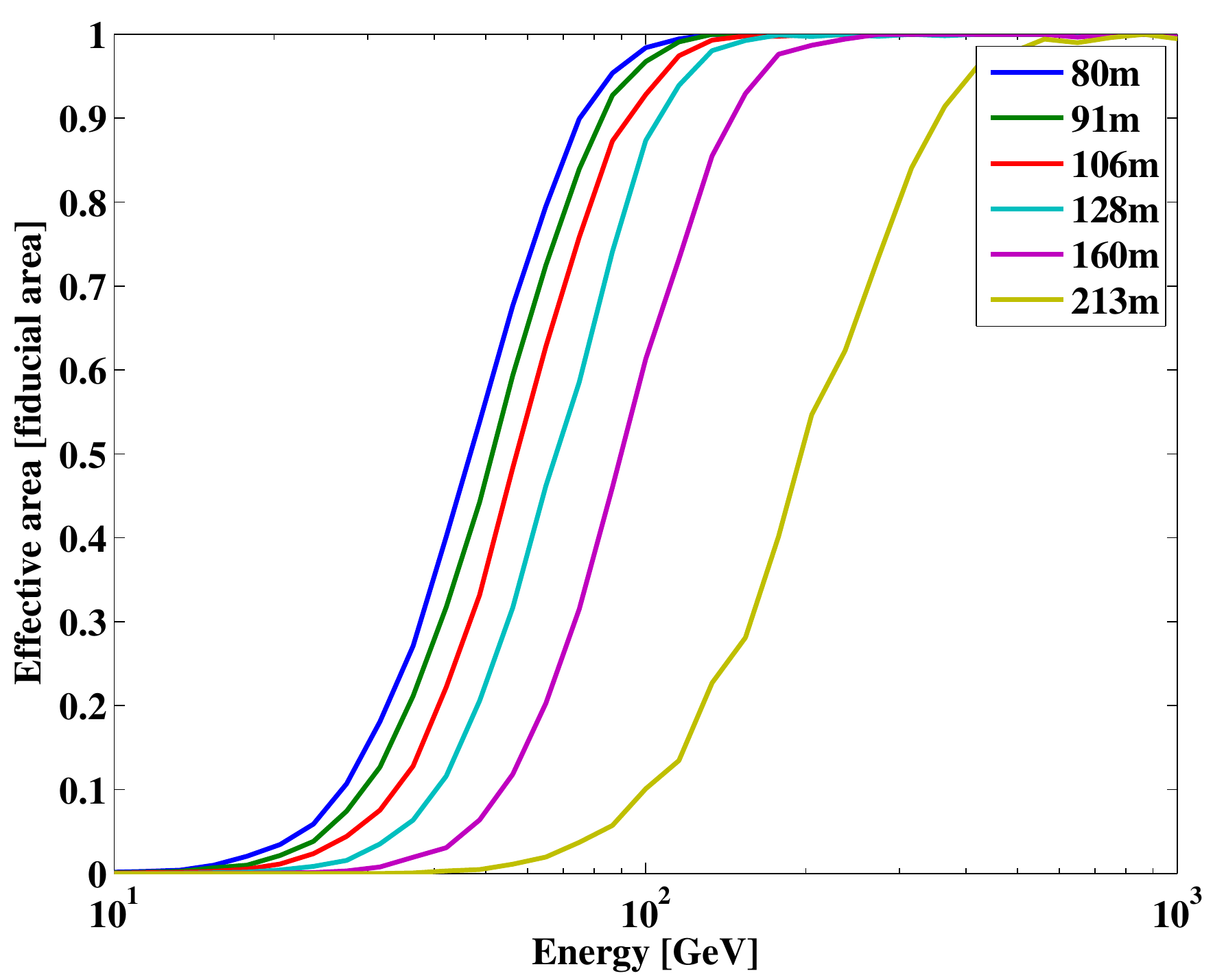}%
\includegraphics[height=0.5\textwidth]{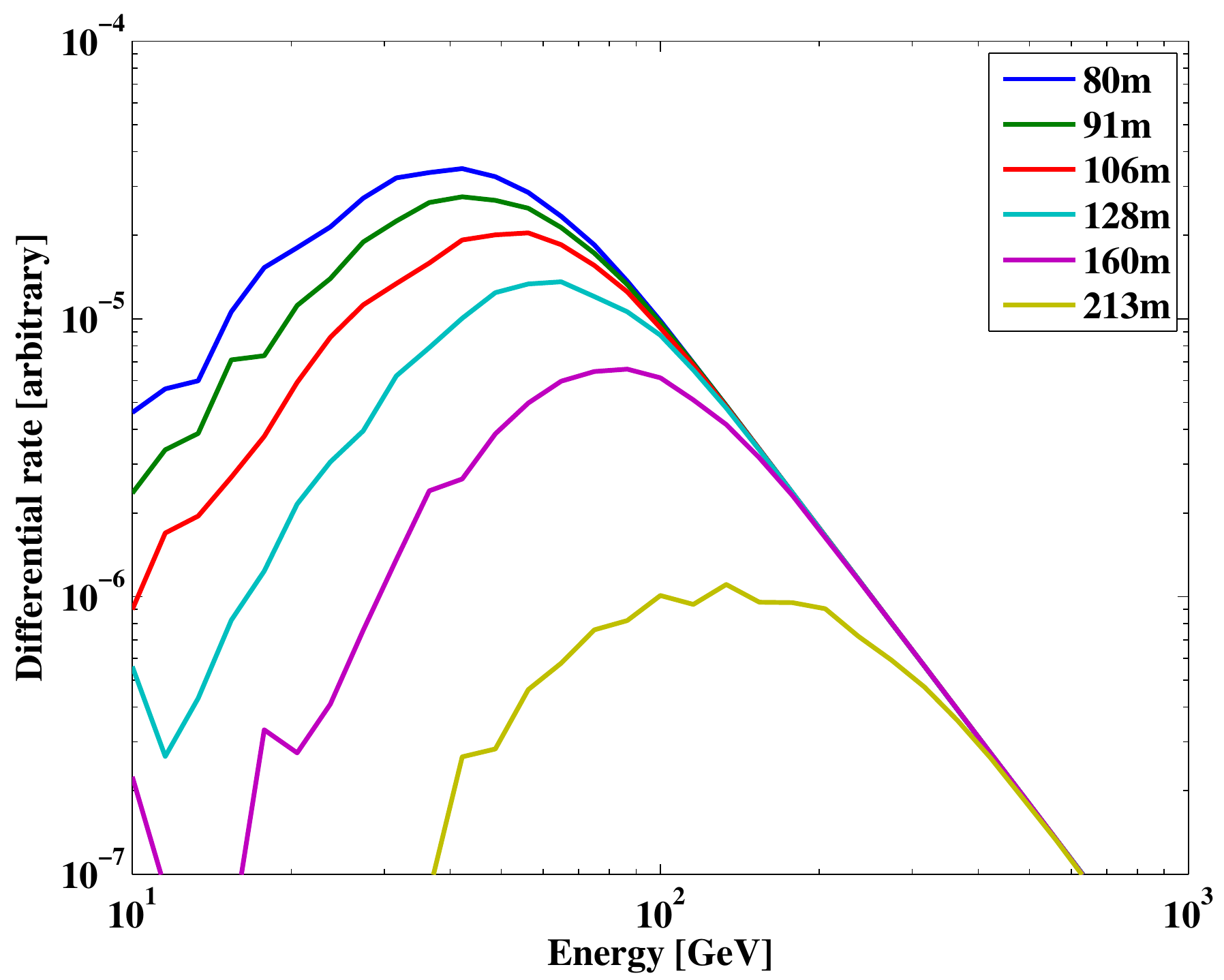}}
\end{center}
\caption{Left: trigger collecting efficiency as a function of energy for 
$\gamma$-ray events impacting a cell in an infinite hexagonal
telescope array, for various telescope separations. Each telescope has
an effective diameter of 7~m and an idealized optical response. Right:
differential rate of detected $\gamma$-rays from a source with a Crab
Nebula-like spectrum.}\label{FIG::SEVEN_METERS}
\end{figure*}

Figure~\ref{FIG::SEVEN_METERS} shows the triggering efficiency as a
function of $\gamma$-ray energy (left), and the differential rate of
$\gamma$-rays from a source with a Crab Nebula-like spectrum (right) for
arrays with $D=7$~m with different telescope separations. In the
highest energy regime, the response of each configuration is
saturated, with the trigger efficiency reaching 1.0 in each case, and
the arrays are indistinguishable in this sense. In a real array,
events impacting near to the perimeter will modify the triggering
efficiency. The significance of this effect is a function of the ratio
of the area of the Cherenkov light pool to the area of the array. It
is likely that, at the highest energies, a $1$~km$^2$ array would have
a collecting area significantly larger than its physical area on the
ground.

At lower energies the triggering efficiency, and hence the collecting
area of the array for $\gamma$-rays, declines. For a VERITAS-like
array the collecting area at low energies is a strong function of
energy due to the exponentially declining density of photons outside
of the central region of the Cherenkov light pool. For large arrays,
in which the telescope separation is smaller than characteristic
Cherenkov diameter, photons are always ``sampled'' within the central
region of the Cherenkov light-pool, in which their density is a weak
function of energy. Due to this ``cell effect'', the fall-off in the
detection rate below the peak energy ($E_\mathrm{peak}$) is
considerably slower than for a VERITAS-like instrument, and this
provides appreciably better sensitivity at low
energies. Figure~\ref{FIG::SEVEN_METERS} (right) shows that for the
$L=80$~m, $D=7$~m configuration, $E_\mathrm{peak}\approx42$~GeV and
the detection rate exceeds 50\% of its peak value within the range
from $\sim18$~GeV to $\sim80$~GeV.

\begin{figure*}[th]
\begin{center}
\includegraphics[width=0.65\textwidth]{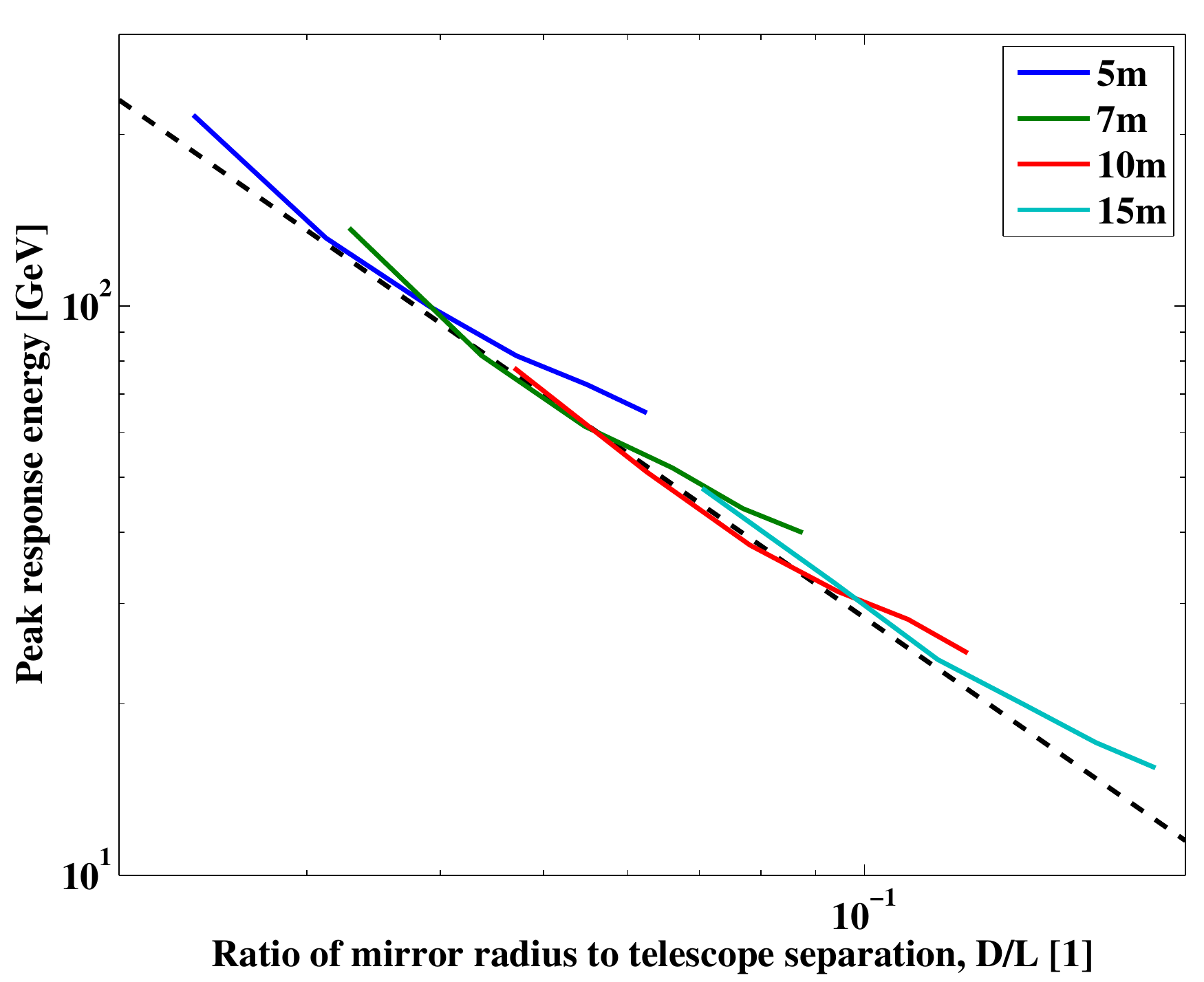}
\end{center}
\caption{Peak response energy, $E_\mathrm{peak}$, of a large array 
as a function of the ratio of the effective mirror diameter, $D$, to
the telescope separation, $L$ (solid curves). The curves lie
approximately on the power-law relation 
$E_\mathrm{peak} = 240\,\mathrm{GeV}
\left(\frac{D/L}{0.02}\right)^{-1.3}$.}\label{FIG::PRE_VS_DS}
\end{figure*}

The results from all the configurations are combined in
figure~\ref{FIG::PRE_VS_DS}, which shows $E_\mathrm{peak}$ as a
function of the ratio of the effective mirror diameter to the
separation between telescopes in the array ($D/L$). An unexpected
finding was that $E_\mathrm{peak}$ is primarily determined by the
single parameter $D/L$, being largely degenerate in the direction of
$D\times L$. This non-trivial result has a direct implication on the
design of future large arrays, which may be required to achieve some
specific $E_\mathrm{peak}$, as determined by the scientific goals of
the experiments. This requirement on $E_\mathrm{peak}$ translates
directly onto a requirement on the ratio of $D/L$,
\[ \left(\frac{D}{7\,\mathrm{m}}\right)\left(\frac{80\,\mathrm{m}}{L}\right) 
= \left(\frac{E_\mathrm{peak}}{36\,\mathrm{GeV}}\right)^{-0.77} \]

\begin{figure*}[th]
\begin{center}
\includegraphics[width=0.65\textwidth]{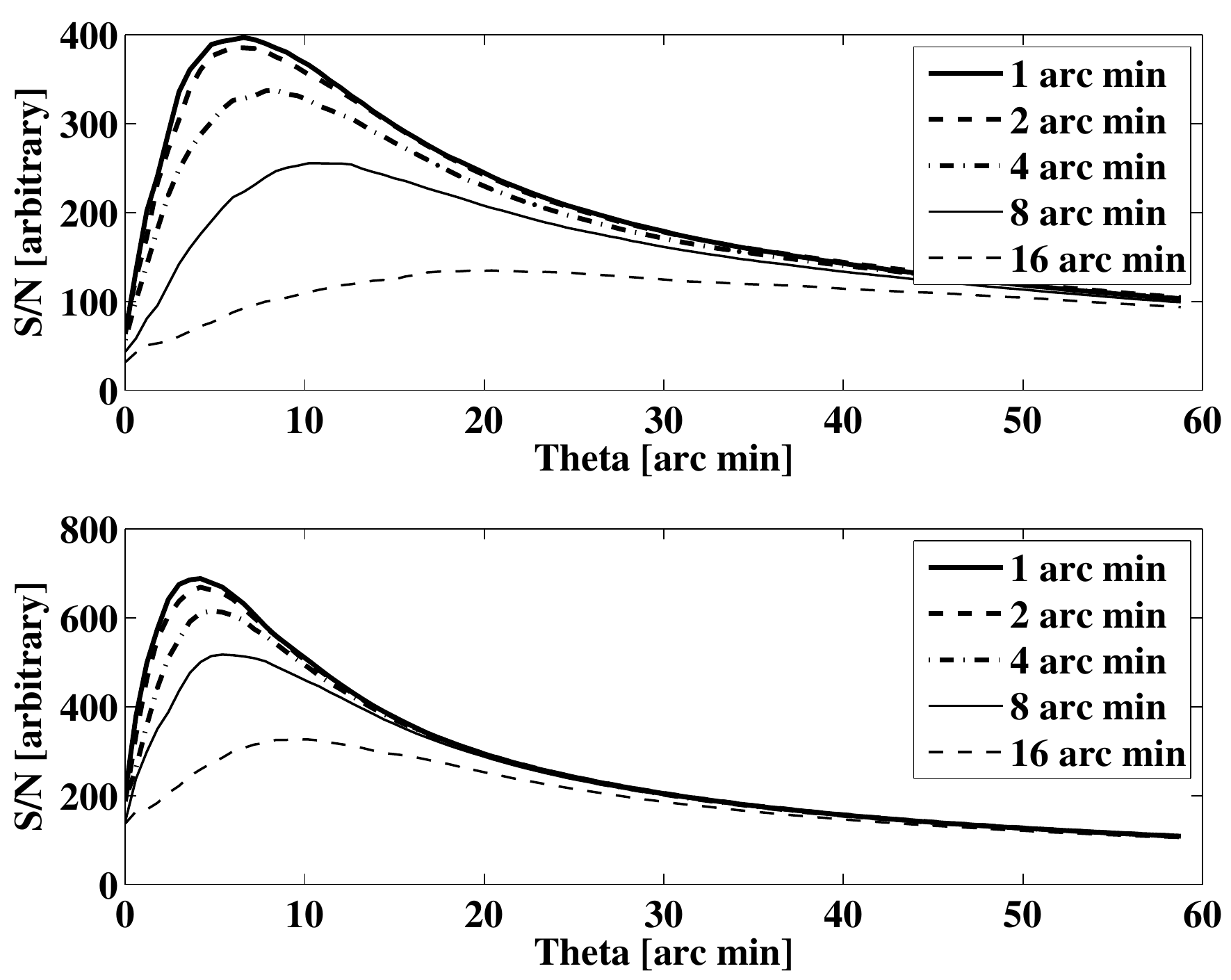}
\end{center}
\caption{Ratio of the number of $\gamma$-ray events reconstructed 
within angular distance $\theta$ of the source to square root of the
number of background events in the same region. Two figures show the
results of simulations for $40$ and $100$~GeV photons.  An array of
$75$~m$^2$ telescopes with separation of $80$~m was
assumed.}\label{FIG::PIXEL_SIZE}
\end{figure*}

Figure~\ref{FIG::PIXEL_SIZE} shows the effects that increasing
pixelation have on the ability of the array to reconstruct the
arrival direction of a primary $\gamma$-ray.  For 1 arc minute pixels
the signal-to-background ratio reaches a maximum at approximately 7
and 3.8 minutes of arc from the source location for 40 and 100~GeV
photons respectively. With 8 arc minute pixels, slightly smaller than
those of current generation instruments such as VERITAS, the ratio
reaches a maximum value at approximately 10 and 5 minutes of arc. The
relative amplitude of the peaks indicate reduction of the instrument
sensitivity by $\sim 40\%$.

Our simulations confirm that the angular resolution of the
reconstructed arrival direction improves with finer pixelation of the
camera until the typical angular scale determined by the transverse
size of the shower core is reached. This size is approximately a few
minutes of arc, as the core, consisting of the highest energy
particles in the cascade, has an extent of a few meters
\cite{REF::VASSILIEV::CHESS_2000}, and the typical observing distance
of 100~GeV cascades is on the order of 10~km.

\section{Conclusions}

A large array of Atmospheric Cherenkov Telescopes will operate in a
very different regime to VERITAS and H.E.S.S. at low energies. The
response is improved largely due to the ``cell effect'', enabling a
significant reduction to the peak detection energy,
$E_\mathrm{peak}$. This parameter, constrained by the scientific goal
of the future instrument, fixes the ratio of $D/L$ for the array,
through a non-trivial scaling law. Decrease of the camera pixel size
to 1 arc minute improves the reconstruction of the photon arrival
direction, and therefore the signal to noise ratio. Considerations of
cost, however, will determine the final design of the future instrument.

\bibliography{icrc0775}
\bibliographystyle{plain}

\end{document}